\documentclass[twoside]{article}
\usepackage{qic,epsfig}

\newcommand{\beq}{\begin{equation}} \newcommand{\eeq}{\end{equation}}
\newcommand{\beqa}{\begin{eqnarray}}
\newcommand{\eeqa}{\end{eqnarray}} \newcommand{\lam}{\lambda}
 \newcommand{\rh}{\rho}
\newcommand{\ga}{\gamma} 
 
 \newcommand{\si}{\sigma}
 \newcommand{\om}{\omega}

\newcommand{\qed}{\nobreak \ifvmode \relax \else
       \ifdim\lastskip<1.5em \hskip-\lastskip
       \hskip1.5em plus0em minus0.5em \fi \nobreak
       \vrule height0.75em width0.5em depth0.25em\fi}

\usepackage{epsfig}
\usepackage{graphicx}
\usepackage{dcolumn}

\def\pra#1{{ Phys.\ Rev. A\/} {\bf#1}} \def\prb#1{{ Phys.\ Rev. B\/} {\bf#1}}
 \def\prl#1{{ Phys.\ Rev.\
Lett.} {\bf#1}}

\begin{document}
\setlength{\textheight}{8.0truein}    

\runninghead{  EVOLUTION FROM ENTANGLEMENT TO DECOHERENCE $\ldots$}
            {T. Yu and J. H. Eberly}

\normalsize\textlineskip
\thispagestyle{empty}
\setcounter{page}{1}

\copyrightheading{0}{0}{2006}{000--000}

\vspace*{0.88truein}

\alphfootnote

\fpage{1}

\centerline{\bf
EVOLUTION FROM ENTANGLEMENT TO DECOHERENCE  }
\vspace*{0.035truein}
\centerline{\bf   OF  BIPARTITE  MIXED ``X"  STATES}
\vspace*{0.37truein}
\centerline{\footnotesize
Ting  Yu\footnote{Email: ting@pas.rochester.edu}    \,\,\,\,   and   J. H. Eberly\footnote{Email: eberly@pas.rochester.edu}}
\vspace*{0.015truein}
\centerline{\footnotesize\it Department of Physics and Astronomy, University
of Rochester}
\baselineskip=10pt
\centerline{\footnotesize\it  Rochester,  NY 14627, USA }

\centerline{\footnotesize 
}

\centerline{\footnotesize\it  }

\centerline{\footnotesize\it }
\vspace*{0.225truein}
\publisher{(received date)}{(revised date)}

\vspace*{0.21truein}




\abstracts{
We  examine a class of bipartite mixed states which we call  X states 
and report
several analytic results related to the occurrence of so-called
entanglement sudden death (ESD) under time evolution in the presence of
common types of environmental noise.   Avoidance of sudden death by
application of purely local operations is shown to be feasible in some cases.
}{}{}

\vspace*{10pt}


\vspace*{3pt}

\communicate{to be filled by the Editorial}

\vspace*{1pt}\textlineskip    







\section{Introduction}
Inevitably, the entanglement of a multi-partite quantum
state becomes degraded with time due to experimental and
environmental noise. The influence of noise on bipartite entanglement
is a problem in the theory of open systems \cite{Yu-Eberly06B}, as
well as of practical importance in any application using quantum
features of information \cite{CM}.

The topic of evolution of quantum coherence in the presence of noise 
sits between two well-investigated problems. One of these is the 
relaxation toward steady-state of one-body coherence of a simple 
quantum system (spin, atom, exciton, quantum dot, etc.) in contact 
with a much larger reservoir \cite{Slichter78}. The other is the 
newer two-body problem where the evolution of the disentanglement of 
the system from its environment is of interest. It is generally 
understood that the latter decoherence occurs much more rapidly than 
the former.

Recently, a practical problem that includes parts of both has drawn 
attention -- the survival of the joint entanglement of two small 
systems with each other while each is exposed to a local noisy 
environment. Their rapid disentanglement from their environments is 
supposed not to be observed, but their disentanglement from each 
other is considered interesting and potentially important.

We have shown in a specific instance of such bipartite 
disentanglement of qubits \cite{Yu-Eberly02,Yu-Eberly03} that 
entanglement is lost in a very different way compared  to the usual 
one-body decoherence measured by the decay of off-diagonal elements 
of the density matrix of either qubit system separately. More 
surprisingly,  we have shown \cite{Yu-Eberly04,Yu-Eberly05} that the 
presence of either pure vacuum noise or even classical noise can 
cause entanglement to decay to zero in a finite time, an effect 
that is labelled ``entanglement sudden death'' (ESD). In the last few years the 
issue of such entanglement decoherence has been discussed in a number 
of distinct contexts such as qubit pairs \cite{Yu-Eberly02, 
Yu-Eberly03, Yu-Eberly04, Lucamarini-etal04, Jakobczyk-Jamroz04, 
Tolkunov-etal05, Ban06, Ban-Shibata06,Malinovsky06, Glendinning-etal, 
An-etal06, Liang06, Roszak-Machnikowski06, Jamroz06, Ficek-Tanas06, 
Solenov05}, finite spin chains \cite{Pratt-Eberly01, Kamta-Starace02, 
WangJ-etal05, Khveshchenko03, Grigorenko-Khveshchenko05, Pratt04a}, 
multipartite systems \cite{Carvalho-etal04, Carvalho-etal05}, 
decoherence dynamics in adiabatic entanglement \cite{Sun01}, 
entanglement transfer \cite{Lamata-etal06}, and open quantum systems 
\cite{Diosi03, Dodd-Halliwell04, Dodd04,Zyczkowski02}, to name a few. 
In addition, a proposal for the direct measurement of finite-time 
disentanglement in a cavity QED context has been made recently 
\cite{Santos-etal06}.

In this paper we report several steps that we expect will assist 
further understanding of this complex and fundamental topic. We focus 
on the smallest and simplest non-trivial situations, in order to help 
expose  consequences that are dynamically fundamental, as opposed to 
ones originating simply in one or another kind of complexity. For 
greatest utility, this more or less mandates that results should be 
analytic rather than numeric. We will treat two qubits prepared in a 
mixed state as an information  time-evolution question 
in the presence of noises. For this purpose, solutions of the 
appropriate Born-Markov-Lindblad master equations can be obtained 
\cite{BMLexamples}  and we will use a Kraus operator approach 
throughout the paper \cite{Kraus}.

The focus will be maintained strictly on the way information itself 
evolves by considering the entanglement of two quantum systems $A$ 
and $B$ exposed to local noises but completely isolated from 
interacting with each other. We will examine evolution toward 
complete disentanglement in a class of commonly occurring bipartite 
density matrices (which we call ``X'' states)  and establish: (a) that 
X-state character is robust, i.e., an X state remains an X state in its 
evolution under the most common noise influences, (b) as Werner 
states are a subclass of X states, we will show that there exisits a 
critical Werner fidelity  below which termination of 
entanglement must occur in a finite time, and (c) that there are 
purely local operations that can sometimes be used to alter the survival 
dynamics of bipartite entanglement. We show that ESD will customarily 
occur, but that in some cases it can be avoided by applying 
appropriate local operations initially.  We  will illustrate all of 
these in the following sections.

The paper is organized as follows: In Sec.  \ref{modelsection},  we 
present two-qubit models where the bipartite system is coupled to 
external sources of phase-damping and amplitude-damping noises. 
Explicit time-dependent solutions in terms of Kraus operators are 
given.  Sec. \ref{concurrence/standard} deals with concurrence, the 
chosen measure of entanglement, and the defining character of an 
X state. In Sec. \ref{decoherence}, the evolution of a Werner state 
toward decoherence is discussed, as an important example of X-state 
behavior under the influence of noise. We find a new fidelity 
boundary below which entanglement sudden death (ESD) must occur for 
all Werner states. In the following Sec. \ref{depolar} we derive the 
ESD that is encountered with depolarizing noise for the X states. In 
Sec. \ref{fragile},  we show that in some cases it is feasible to 
transform a short-lived  state into a long-lived state by applying 
specified local operations initially, and we conclude in Sec. 
\ref{conc}.

\section{Models}
\label{modelsection}
\noindent
The non-interacting quantum systems $A$ and $B$  and their separate 
reservoirs labeled $a$ and $b$ are assumed to follow the same evolution route
separately. We use the familiar Hamiltonian (for qubit $A$ say):
\beq
\label{model}
H^A_{\rm tot}= H^A_{\rm at} + H^a_{\rm res} + H^{\rm Aa}_{\rm int},
\eeq
where:
\beq
H^A_{\rm at} =
\frac{1}{2}\om_A \si^A_z  \quad {\rm and} \quad
H^a_{\rm res} = \sum_{{k}}\om_{k}a_{ k}^\dag a_{ k}
\label{eq1}
\eeq
and for exposure to phase and amplitude noises the interaction Hamiltonians
$H_{\rm int}$ are given by
\beq
H^{Aa}_{\rm ph-int} = \sum_{{k}}  \si^A_z (f_{ k}a^\dag_{k}
+ f_{ k}^*a_{ k}),
\label{eq3}
\eeq
and
\beq
\label{intam}
H^{Aa}_{\rm am-int} = \sum_{{ k}} ( g_{ k} \si^A_- a^\dag_{k}
+ g_{k}^* \sigma^A_+a_{ k}).
\eeq

Here the $a_{ k}$ are bosonic reservoir coordinates satisfying $[a_{ 
k}, a_{ k'}]=\delta_{ k, k'},$ the $g_{ k}$ are  broadband coupling 
constants, and the $\si^A$s denote the usual Pauli matrices for qubit
$A$. The same forms hold for $B$, with a set of bosonic reservoir 
coordinates $b_k$.  $\{A,a\}$ and $\{B, b\}$ are 
completely independent,  and no 
decoherence-free joint subspaces are available.  As remarked, these 
total Hamiltonians provide well-known solvable qubit-reservoir 
interactions, but we are interested in the evolution of joint 
information as a consequence of the completely separate interactions.

We consider qubits $A$ and $B$ prepared in a mixed state. For this purpose, solutions of the 
appropriate Born-Markov-Lindblad equations can be obtained via 
several routes, and we find the Kraus operator form \cite{Kraus} 
convenient for our purpose. Given an initial state $\rho$ (pure or 
mixed) for two qubits $A$ and $B$,  its evolution can be written 
compactly as
\beq \label{Kraus}
\rho(t) = \sum_\mu K_\mu(t)\rho(0) K_\mu^\dag(t),
\eeq
where the so-called Kraus operators $K_\mu$ satisfy $\sum_\mu
K_\mu^\dag K_\mu = 1$ for all $t$. Obviously, the Kraus operators
contain the complete information about the system's dynamics.

In the case of  dephasing noise one has following compact Kraus  operators:
\begin{eqnarray}
         \label{k1}E_1
&=&\left(\begin{array}{clcr}
\gamma_A && 0\\
0 && 1\\
\end{array}
          \right)\otimes  \left(
\begin{array}{clcr}
\ga_B & 0\\
0 & 1\\
\end{array}
          \right),\\
          E_2&=&\left(\begin{array}{clcr}
\gamma_A && 0\\
0 && 1\\
\end{array}
          \right)\otimes\left(
\begin{array}{clcr}
0 & 0 \\
0 & \om_B\\
\end{array}
          \right),\\
		E_3&=& \left(
\begin{array}{clcr}
0 & 0\\
0 & \om_A \\
\end{array}
          \right) \otimes \left(
\begin{array}{clcr}
\ga_B & 0\\
0 & 1\\
\end{array}
          \right),\\
		E_4 &=& \left(
\begin{array}{clcr}
0 & 0\\
0 & \om_A\\
\end{array}
          \right)\otimes \left(
\begin{array}{clcr}
0 &  0 \\
0 & \om_B\\
\end{array}
          \right),\label{k5}
          \end{eqnarray}
where the time-dependent Kraus matrix elements are $$\gamma_A(t) = 
\exp{(-\Gamma^A_{\rm ph} t/2)} \quad {\rm and} \quad \om_A(t) = 
\sqrt{1-\gamma^2_A(t)},$$ where $\Gamma^A_{\rm ph}$ is the phase 
damping  rate of qubit A.  We use the similar expressions 
$\gamma_B(t)$ and $\om_B(t)$ for qubit B,  and will take 
$\Gamma^A_{\rm ph} = \Gamma^B_{\rm ph}=\Gamma_{\rm ph}$ for greatest 
simplicity.

Similarly, the Kraus operators for zero-temperature amplitude noise 
are given by
\begin{eqnarray}
       \label{k10}
F_1&=&\left(\begin{array}{clcr}
\gamma_A & 0\\
0 &  1\\
\end{array}
        \right)\otimes\left(
\begin{array}{clcr}
\gamma_B & 0 \\
0 &  1\\
\end{array}
          \right),\label{e1}\\
        F_2&=&\left(
\begin{array}{clcr}
\gamma_A   &   0 \\
0  & 1\\
\end{array}
        \right)\otimes\left(
\begin{array}{clcr}
0 & 0 \\
\omega_B  & 0\\
\end{array}
          \right),\label{e2}\\
F_3&=& \left(\begin{array}{clcr}
0 &   0 \\
\om_A  &  0\\
\end{array}
        \right)\otimes\left(
\begin{array}{clcr}
\gamma_B & 0 \\
0 & 1\\
\end{array}
          \right),\label{e3}\\
F_4 &=&  \left(\begin{array}{clcr}
0  &  0) \\

\omega_A &   0 \\
\end{array}
        \right)\otimes\left(
\begin{array}{clcr}
0 & 0 \\
\omega_B & 0 \\
\end{array}
          \right),\label{e4}
        \end{eqnarray}
and the time-dependent Kraus matrix elements are defined similarly as in
the dephasing model, e.g.,  $\gamma_A(t)=\exp \left(-\Gamma^A_{\rm am 
} t/2\right)$, etc. With the above explicit solutions of the models, 
we are able to compute the degree of entanglement of the two qubits 
in temporal evolution.

\section{Decoherence Measure and X States}
\label{concurrence/standard}

In order to describe the dynamic evolution of quantum entanglement we 
use Wootters' concurrence \cite{Wootters}.  Any entropy-based measure 
of entanglement will yield the same conclusion about bipartite 
separability.  Concurrence varies from $C=0$ for a separable state to 
$C=1$ for a maximally entangled state. For any two qubits, the 
concurrence may be calculated explicitly from their density matrix 
$\rho$ for qubits $A$ and $B$:

\beq
\label{definationc}
C(\rh)=\max\{0,\sqrt{\lam_1}-\sqrt{\lam_2}-\sqrt{\lam_3}-\sqrt{\lam_4}\,\,\},
\eeq
where the quantities $\lam_i$ are the eigenvalues in decreasing order 
of the matrix $\zeta$:
\beq \zeta=\rho(\sigma_y\otimes
\sigma_y)\rho^*(\sigma_y\otimes \sigma_y),
\label{concurrence}
\eeq
where $\rh^*$ denotes the complex
conjugation of $\rh$ in the standard basis $|+ +\rangle, |+ -\rangle, 
|- +\rangle, |- -\rangle$ and
$\si_y$ is the Pauli matrix expressed in the same basis as:
\begin{equation}
\si_y= \left(\begin{array}{clcr}
0  &  -i\\
i  &   0 \\
\end{array}
        \right).
\end{equation}

In the following we will examine the evolution of entanglement under 
noise-induced relaxation of a class of important  bipartite density 
matrices which are defined below. Since a density matrix in this 
class  only contains non-zero elements in an ``X" formation, along 
the main diagonal and anti-diagonal, we call them ``X states":
\begin{equation}
  \label{e.oldrho}
\rho^{AB} =\left(
\begin{array}{clcr}
a & 0  &  0 & w\\
0  & b & z & 0 \\
0  & z^* & c & 0\\
w^*  &  0 & 0 & d
\end{array} \right).
\end{equation}
where $a+b+c+d = 1$.

Such a simple matrix is actually not unusual. Experience shows that
this    X  mixed state arises naturally in a wide variety of
physical situations (see \cite{WangJ-etal05,Pratt04a,standard2}). We
particularly note that it includes pure Bell states as well as the
well-known Werner mixed state \cite{Werner} as special cases. Unitary
transforms of it extend its domain even more widely, as we will
explain below.

The mixed states defined here not only are rather common but also 
have the property that they often retain the X form under noise 
evolution. This may be expected for phase noise, which can only give 
time dependence to the off-diagonal matrix elements. The interaction 
Hamiltonian and Kraus operators for  amplitude (e.g., 
quantum vacuum) noise evolution are different, and evolution under 
amplitude noise is more elaborate, affecting all six non-zero 
elements (see \cite{Yu-Eberly04}), but robust form-invariance during 
evolution is easy to check. This very simple finding applies to a 
wide array of realistic noise sources.

For the X state defined in (\ref{e.oldrho}),  concurrence 
\cite{Wootters} can be easily computed as
$$C(\rho^{AB})=2\max\{0, |z|-\sqrt{ad}, |w|-\sqrt{bc}\}.$$

\section{Evolution to Decoherence of the Werner State}
\label{decoherence}
\noindent
Now we examine decoherence evolution under first phase damping 
and then amplitude damping.  Within the set of X matrices, let us 
focus now on a Werner state \cite{Werner,curious}:
\beq \label{werner}
\rho_W=\frac{1-F}{3}I_4 +
\frac{4F-1}{3}|\Psi^-\rangle\langle \Psi^-|,
\eeq
whose matrix elements can be matched to those of the X state 
$\rho^{AB}$ easily.  $F$ is termed fidelity, and $1 \ge F \ge 
\frac{1}{4}$. We will begin by obtaining the time-dependence of 
entanglement for the Werner state.  Under phase noise the only time 
dependence is in $z$:
\beq \label{zphase}
z(t) = \frac{1-4F}{6} \gamma^2(t), \quad {\rm with} \quad \gamma(t)
\equiv e^{-\Gamma_{\rm ph}t/2}.
\eeq

\begin{figure} [htbp]

\vspace*{13pt}

\centerline{\epsfig{file=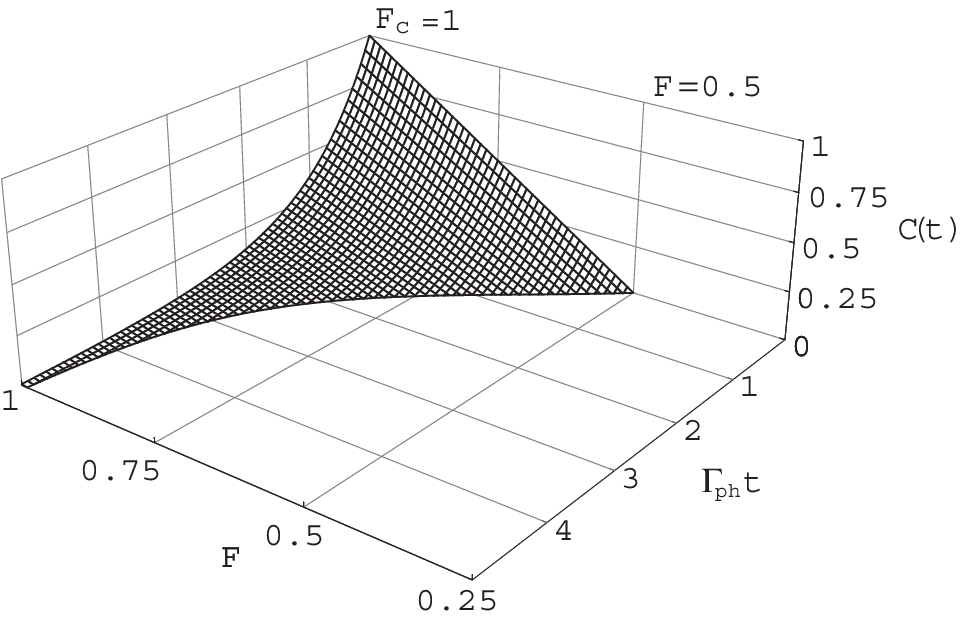, width=8.2cm}} 

\vspace*{13pt}

\fcaption{\label{fig2}  Phase noise causes $\rho_W$ to
disentangle completely in finite time for all Werner states except in
the limiting case of a pure Bell state. The graph shows  $C(t)$ vs. $F$
and $\Gamma_{\rm ph} t$.}

\end{figure}

The results are shown in Fig. \ref{fig2}. In particular we note the 
occurrence of ESD,  in which concurrence non-smoothly goes to zero at 
a finite time (and remains zero). This is apparent for all initial 
$F<1$. It has been noted already for quantum vacuum noise qubit 
decoherence \cite{Yu-Eberly04} and for disentanglement of continuous 
joint states \cite{Diosi03, Dodd-Halliwell04, Dodd04}. The analytic 
expressions above make it clear why this is so. Since the matrix 
elements $a$ and $d$ are fixed, as $z$ decays it must become less 
than $\sqrt{ad}$ at a specific time $\tau_{\rm ph}$, which can be 
easily determined to be given by
\beq \label{phaseratio}
\frac{\tau^{\rm ph}}{\tau_0} = {\ln} 
\left[\frac{4F-1}{2-2F}\right],\,\,\,\,  1>F>\frac{1}{2}
\eeq
where $\tau_0 = 1/\Gamma_{\rm ph}$ marks the $1/e$  point in the purely
exponential decay of the underlying individual qubits.

Next we consider Werner state evolution under amplitude noise, and we
find from the appropriate Kraus operators given in 
(\ref{k10}-\ref{e4})  \cite{Yu-Eberly04}
that the following time dependences specify $\rho_W(t)$ at any time:
\beq
z(t)=\frac{1-4F}{6} \ga^2,
\eeq
\beq
a(t)=\frac{1-F}{3} \ga^4,
\eeq
\beq
b(t)=\frac{2F+1}{6} \ga^2 +\frac{1-F}{3}\ga^2\om^2,
\eeq
\beq
c(t)=\frac{2F+1}{6} \ga^2 +\frac{1-F}{3}\ga^2\om^2, \quad {\rm and}
\eeq
\beq
d(t)=\frac{1-F}{3} +\frac{2F+1}{3}\om^2 + \frac{1-F}{3}\om^4.
\eeq
In principle the time-dependent $\gamma$ and $\omega$ parameters 
could be different for qubits $A$ and $B$, but we again take them 
identical and write $\ga = \exp[-\Gamma_{\rm am} t/2]$ and $\om^2 = 1 
- \ga^2$, where we use $\Gamma_{\rm am}$ to denote the upper level 
decay rate of the qubits.

\begin{figure} [htbp]

\vspace*{13pt}

\centerline{\epsfig{file=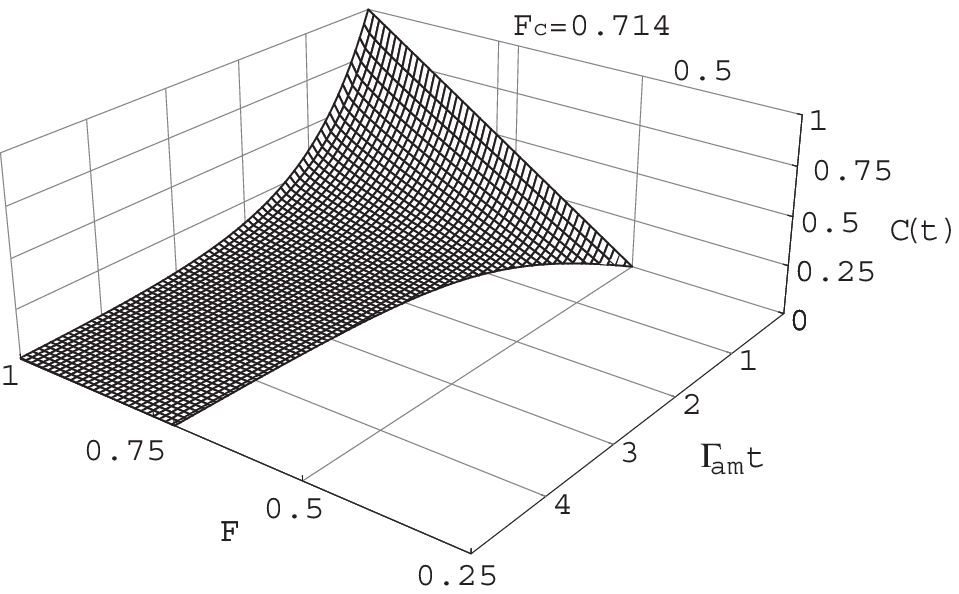, width=8.2cm}} 

\vspace*{13pt}

\fcaption{\label{fig1}  In the presence of amplitude
noise there is long-lived concurrence of Werner states only  for
sufficiently high fidelity, $F > F_{\rm c} \simeq 0.714$. The graph
shows the critical fidelity boundary in plotting $C(t)$ vs.
$\Gamma_{\rm am} t$.}

\end{figure}

Sudden death of Werner state entanglement appears here also, but with the 
important added element that sudden death from amplitude noise occurs 
only for a low range of fidelity values. Our result shows that for 
all initial $F$ above the critical fidelity  $F_{\rm c} \simeq 0.714$ 
entanglement remains finite for all time, and has an infinitely long 
smooth decay, faster than but similar to, the decay of single-qubit 
coherence. This is shown in the plot in Fig.~\ref{fig1}.

\section{Bistability Decoherence}
\label{depolar}
\noindent
In this section, we discuss the entanglement decoherence of a 
representative X matrix under bistability noise, by which we mean 
noise that induces incoherent random transfer back and forth between 
the two qubit states when they are energetically degenerate. Examples 
occur in bistable systems of all kinds, for example in semiconductor 
junctions or double-well electron potentials in photonic crystals. A 
physically different example is polarization of photons in optical 
fiber with indeterminately random local birefringence.  In all of 
these cases we can speak of the effect as arising from exposure to an 
infinite-temperature reservoir. In that case the 
population-equalizing up-transfer and down-transfer rates will be 
denoted $\Gamma_{\rm eq}$ and taken the same for the two qubits. 
Given these remarks, our basic model with Hamiltonians (\ref{model}) 
and (\ref{intam}) still allows a useful Kraus representation and the 
Kraus matrices are given by:
\begin{eqnarray}
G_1&=& \frac{1}{\sqrt 2}\left(
\begin{array}{cc}
\gamma(t) & 0\\
0 &  1\\
\end{array}
        \right), \label{G1}\\
G_2 &=& \frac{1}{\sqrt 2} \left(
\begin{array}{cc}
0 & 0 \\
\om(t)  & 0\\
\end{array}
          \right), \label{G2}\\
G_3 &=& \frac{1}{\sqrt 2} \left(
\begin{array}{cc}
1 &   0 \\
0  &  \gamma(t)\\
\end{array}
        \right), \label{G3}\\
G_4 &=& \frac{1}{\sqrt 2}  \left(
\begin{array}{cc}
0  &  \omega(t) \\
0  &  0 \\
\end{array}
        \right), \label{G4}
\end{eqnarray}
where now
$$\gamma(t) = \exp{(-\Gamma_{\rm eq} t/2)} \quad {\rm and} \quad
\om(t) = \sqrt{1-\gamma^2(t)}.$$
State equalization presents the extreme opposite case from vacuum
noise, in the sense that strong equalization treats both qubit states
equally incoherently, whereas vacuum noise induces incoherent decay
into just the energetically lower of the two states.

One finds that the X form of the  density matrix (\ref{e.oldrho})  is
still preserved under state equalizing noise and so at time $t$ it 
retains the X form (we set $w=0$ for simplicity):
\begin{equation}
\label{sol} \rho(t) = \left(
\begin{array}{clcr}
a(t) & 0  &  0 & 0 \\
0  & b(t) & z(t) & 0 \\
0  & z(t) & c(t) & 0\\
0  &  0 & 0 & d(t)
\end{array} \right).
\end{equation}
We assume that the two qubits are affected by two identical local
depolarization noises, and in this case the time-dependent matrix
elements are given by the following:
\beqa
4a(t) &=& \ga^4a + a+\om^2 (b +  c) + \om^4d\nonumber\\
&&+ 2\ga^2a + \ga^2\om^2( b+c),\\
4b(t) &=& 2\ga^2b +\ga^2\om^2 (a+d) \nonumber\\
&&+ b+\ga^4 b+\om^2(a+d) + \om^4c,\\
4c(t) &=& 2\ga^2c +\ga^2\om^2 (a +d) \nonumber\\
&&+ c+\om^2(d + a) + \om^4 b+\ga^4 c,\\
4d(t) &=& d+\om^2(b +  c) + \om^4 a + \ga^4 d\nonumber\\
&&+ 2\ga^2 d + \ga^2\om^2(b+c),\\
z(t) &=& \ga^2 z.
\eeqa
Some algebraic examination shows that this result also leads to ESD,
i.e., bistable equalization leads all entangled X states 
(\ref{e.oldrho}) to become separable states in a finite time.

It is easy to check that when $t \to \infty$,
\beq
z \to z(\infty) = 0,
\eeq
\beq
\{a, b, c, d\}  \to  \{\frac{1}{4}, \frac{1}{4},  \frac{1}{4},\frac{1}{4}\},
\eeq
and the asymptotic separability arising from the equivalence of all
diagonal elements is just a special case of the general theorem by
Zyczkowski,  et al.\cite{Lewenstein98}.

\section{Fragile and Robust Initial Entangled States}
\label{fragile}
\noindent
It is known that entangled states evolve differently under different 
environmental noise influences if special symmetries exist. For 
example, in the case of collective dephasing noise (see, e.g., 
\cite{Yu-Eberly02}), there may exist decoherence-free subspaces in 
which the entangled states are well protected against interaction 
with the noise. For the models presented here the noises influence 
each qubit independently, so there are no decoherence-free subspaces 
and there is no such protection from ESD available. However, we now 
show that it is still possible to avoid sudden death by using 
appropriate local initial preparations.  To illustrate this, we 
consider another mixed state within the category of the X matrix 
defined in (\ref{e.oldrho}):
\beq \label{tildewerner}
\tilde\rho_W=\frac{1-F}{3}I_4 +
\frac{4F-1}{3}|\Phi^-\rangle\langle \Phi^-|,
\eeq
where $|\Phi^-\rangle = (|++\rangle - |--\rangle)/\sqrt{2},$
is a Bell state. In matrix form at any $t>0$ we can write
\begin{equation} \label{tilderho}
\tilde\rho_W(t)=
\left( \begin{array}{clcr}
a(t) & 0  &  0 & w(t) \\
0  & b(t) & 0 & 0 \\
0  & 0 & c(t) & 0\\
w^*(t) & 0 & 0 & d(t)
\end{array} \right).
\end{equation}
It is easy to compute the concurrence of this mixed state:  $C = 2 
\max\{0, |w|-\sqrt{bc}\}$.

Consider the time dependences for $\tilde\rho_W$ obtained from the
amplitude-noise Kraus operators as before. It is easy to check that the
sudden death condition for $\tilde\rho_W$'s concurrence is now:
\beq \label{ConcurtildeW}
\frac{4F-1}{6}\gamma^2  \  =\   \frac{1-F}{3}\gamma^2 +
\frac{2F+1}{6}\gamma^2 \omega^2 ,
\eeq
which is satisfied at a finite $t$ for any value of fidelity $F$.
That is, here there is no range of ``protected" fidelity values under
amplitude noise, as was the case for the other form of X state
in Fig. ~\ref{fig1}. Indefinite survival is impossible in this case,
similar to what the plot in Fig.~\ref{fig2} shows for phase noise.

However, this result has important implications related to survival. 
One easily shows that $\tilde\rho_W$ is closely related to the 
earlier Werner state $\rho_W$, which does have a range of protected 
fidelity values. In fact $\rho_W$ and $\tilde\rho_W$ are unitary 
transforms of each other under a {\em purely local} transformation 
operator: $U = i\sigma_x^A \otimes I_B$.  This shows that survival 
against noise of initial mixed state entanglement can in a wide range 
of situations be dramatically improved by a simple local unitary 
operation (here changing $\tilde\rho_W$ into $\rho_W$), even while 
the degree of entanglement is not changed.  Intuitively, it is easy 
to see that the noise influence represented by the Kraus operators 
varies for different matrix elements of a bipartite density matrix. 
Although local operations cannot change the degree of entanglememt, 
it is possible that   local operations can rearrange the matrix 
elements of the X states such that the resulting density matrix is 
more robust (or fragile) than the original one. Therefore,  ESD may 
be manipulated  by preparatory transformation that is purely local.
\section{Concluding Remarks}
\label{conc}
\noindent
In summary, we have examined quantitatively via fully analytic 
expressions the non-local decoherence properties  of a wide range of 
mixed states. We have used relatively simple Kraus  operators to do 
this. We have also shown that three well understood physical noise 
types (phase noise, amplitude noise, and state-equalizing noise) do 
not alter the form of the X mixed state during evolution, although 
entanglement survival may be long or short. In particular, we  have 
established that Werner states are subject to the sudden death 
effect, and have specified a new critical fidelity boundary below 
which sudden death must occur.  Surprisingly, we have found that 
Werner states are more robust under pure  amplitude noise (e.g., 
spontaneous emission) than under pure phase noise even though 
amplitude noise is in a sense more disruptive than dephasing, as the 
former causes diagonal and off-diagonal relaxation and the latter 
off-diagonal relaxation alone.  Moreover,  we have shown that 
bistability decoherence can cause  all the X states to disentangle 
completely in finite time. In addition, we have shown that in some 
cases an initial mixed state's entanglement can be preserved under 
subsequent noisy evolution, i.e., sudden death avoided, by an initial 
local unitary operation. Such local operations may offer a useful 
tool in entanglement control when the duration of entangled states is 
crucial in the processes of quantum state storage and preparation.

Finally,  we note that while we have found interesting and unexpected 
features of Werner and other X mixed states in the presence of common 
noise sources, our Kraus operators treated all of them as white 
(Markovian) noises. It will be an important theoretical challenge to extend these 
results to the case of non-Markovian environmental influences 
\cite{Glendinning-etal,Yonac-etal06,Hu-etal00-05}.

\nonumsection{Acknowledgements}

\noindent

We acknowledge financial support from NSF Grant
PHY-0456952 and ARO Grant W911NF-05-1-0543.
We also thank C. Broadbent for assistance with Figs. 1  and 2.

\nonumsection{References}

\end{document}